\documentclass[sigconf]{acmart}

\AtBeginDocument{%
  }

\copyrightyear{2024}
\acmYear{2024}

\copyrightyear{2024}
\acmYear{2024}
\setcopyright{rightsretained}
\acmConference[L@S '24]{Proceedings of the Eleventh ACM Conference on Learning @ Scale}{July 18--20, 2024}{Atlanta, GA, USA}
\acmBooktitle{Proceedings of the Eleventh ACM Conference on Learning @ Scale (L@S '24), July 18--20, 2024, Atlanta, GA, USA}
\acmDOI{10.1145/3657604.3664695}
\acmISBN{979-8-4007-0633-2/24/07}

\makeatletter
\gdef\@copyrightpermission{
 \begin{minipage}{0.3\columnwidth}
 \href{https://creativecommons.org/licenses/by/4.0/}{\includegraphics[width=0.90\textwidth]{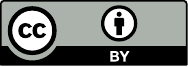}}
 \end{minipage}\hfill
 \begin{minipage}{0.7\columnwidth}
 \href{https://creativecommons.org/licenses/by/4.0/}{This work is licensed under a Creative Commons
Attribution International 4.0 License.}
 \end{minipage}
 \vspace{5pt}
}
\makeatother

\usepackage{enumitem}
\usepackage{balance} 
\begin{document}

\title{SimPal: Towards a Meta-Conversational Framework to Understand Teacher's Instructional Goals for K-12 Physics}


\author{Effat Farhana}
\email{effat.farhana@vanderbilt.edu}
\affiliation{%
  \institution{Vanderbilt University}
  \city{Nashville, Tennessee}
  \country{USA}}

\author{Souvika Sarkar}
\email{szs0239@auburn.edu}
\affiliation{%
  \institution{Auburn University}
  \city{Auburn, Alabama}
  \country{USA}}

\author{Ralph Knipper}
\email{rak0035@auburn.edu}
\affiliation{%
  \institution{Auburn University}
  \city{Auburn, Alabama}
  \country{USA}}

\author{Indrani Dey}
\email{idey2@wisc.edu}
\affiliation{%
  \institution{University of Wisconsin-Madison}
  \city{Madison, Wisconsin}
  \country{USA}}

\author{Hari Narayanan}
\email{naraynh@auburn.edu}
\affiliation{%
  \institution{Auburn University}
  \city{Auburn, Alabama}
  \country{USA}}

\author{Sadhana Puntambekar}
\email{puntambekar@education.wisc.edu}
\affiliation{%
  \institution{University of Wisconsin-Madison}
  \city{Madison, Wisconsin}
  \country{USA}}

\author{Santu Karmaker}
\email{sks0086@auburn.edu}
\affiliation{%
  \institution{Auburn University}
  \city{Auburn, Alabama}
  \country{USA}}

\renewcommand{\shortauthors}{Effat Farhana et al.}

\begin{abstract}

Simulations are widely used to teach science in grade schools. These simulations are often augmented with a conversational artificial intelligence (AI) agent to provide real-time scaffolding support for students conducting experiments using the simulations. AI agents are highly tailored for each simulation,  with a predesigned set of Instructional Goals (IGs), making it difficult for teachers to adjust IGs as the agent may no longer align with the revised IGs. Additionally, teachers are hesitant to adopt new third-party simulations for the same reasons. In this research, we introduce SimPal, a Large Language Model (LLM) based meta-conversational agent, to solve this misalignment issue between a pre-trained conversational AI agent and the constantly evolving pedagogy of instructors. Through natural conversation with SimPal, teachers first explain their desired IGs, based on which SimPal identifies a set of relevant physical variables and their relationships to create symbolic representations of the desired IGs. The symbolic representations can then be leveraged to design prompts for the original AI agent to yield better alignment with the desired IGs. We empirically evaluated SimPal using two LLMs, ChatGPT-3.5 and PaLM 2, on 63 Physics simulations from PhET and Golabz. Additionally, we examined the impact of different prompting techniques on LLM's performance by utilizing the TELeR taxonomy to identify relevant physical variables for the IGs. Our findings showed that SimPal can do this task with a high degree of accuracy when provided with a well-defined prompt. 
\end{abstract}




\maketitle

\section{Introduction}

Simulations are widely used in science education, and prior research shows that using simulations in science education can enhance students' comprehension of scientific concepts~\cite{Simulations,simulations2}. However, students often need guidance and scaffolding when conducting experiments with simulations \cite{StudentStruggleSimulation2, gonzalez2003convenience}, and it is challenging for one teacher to provide real-time support to multiple students simultaneously \cite{StudentStruggleSimulation}. Recent advancements in Large Language Models (LLMs) \cite{gpt3} have revolutionized conversational AI agents as a plausible solution to provide real-time support to students. But LLM-powered conversational AI agents also present unique challenges. First, existing AI agents are highly customized for a specific simulation with a predesigned set of Instructional Goals (IGs) \cite{TeacherSimulationStruggle}. Therefore, teachers often struggle to edit these predesigned IGs or redesign the IGs because the AI agent will no longer be aligned with the revised IGs. Second, middle or high school science teachers lack the technical expertise to customize AI agents \cite{park2023integrating}. This leads to the use of pre-existing, non-customizable agents or third-party software, which requires more time and resources for simulations. For similar reasons, teachers also hesitate to integrate new/other third-party (closed-source) simulations into their instructional materials.

How can we empower teachers to integrate any third-party (open or closed-source) simulation into their instruction materials such that they can I) freely design their own Instructional Goals (IGs) and II) quickly customize a conversational AI agent to better align with their IGs? More importantly, how can we achieve this goal without requiring teachers to understand the technical details of Large Language Models (LLMs) like GPT-4~\cite{gpt4} and PaLM~\cite{chowdhery2023palm, anil2023palm2}? While LLMs are trained on vast internet text data and can aid in language comprehension tasks like answering questions \cite{LLMQgen} and facilitating human conversations \cite{LLMConversation}, adapting LLMs to domain-specific tasks is still challenging due to a lack of proper knowledge grounding in that particular domain. It is also unrealistic to expect school teachers to learn knowledge-grounding techniques that require in-depth machine learning or deep learning knowledge.

This paper introduces SimPal, a meta-conversational agent that can assist school teachers in adopting any existing physics simulation into their lesson plan while allowing them to custom-design their own IGs and customize a general-purpose LLM that aligns with those custom IGs, facilitating \textit{instruction at scale}. SimPal achieves this ambitious goal through \textit{meta-conversation}, which is essentially a conversation with the teacher about structuring future conversations with students for simulation-based physics experiments. Through natural (meta-)conversation with SimPal, teachers first explain their desired IGs, based on which SimPal identifies a set of relevant physical variables and their relationships to create symbolic representations of the desired IGs. The symbolic representations can then be leveraged to design prompts for the original AI agent to yield better alignment with the desired IGs. 

\begin{figure}[!htb]
 
  \includegraphics[width = \linewidth, height = 0.6\linewidth]{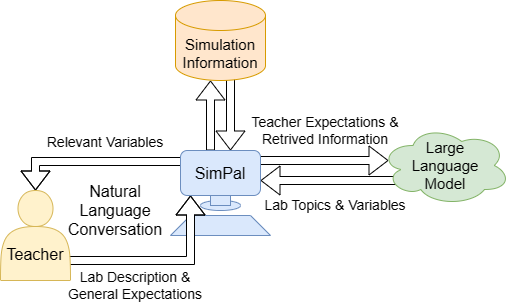}\caption{SimPal's high-level overview: The teacher converses with SimPal, discussing their simulation of interest and corresponding IG. As the conversation progresses, SimPal extracts useful information from the conversation to infer a computational representation of the teacher's IG. That internal representation is then communicated back to the teacher so they can make any necessary adjustments.} 
  \label{fig:Intro}
\end{figure}

Figure~\ref{fig:Intro} presents an overview of SimPal's interaction with the teacher. The teacher conveys their IGs to SimPal, and then SimPal creates symbolic representations of IGs by identifying relevant physical characteristics and their interactions. Accurately identifying relevant physical variables is crucial, as the IGs are encoded in terms of these variables and will guide student interactions. SimPal's architecture allows a teacher to tailor their lesson plan by I) modifying the variables and relations of a simulation through natural conversation and II) integrating any third-party simulation. 

%


A challenging first step toward achieving this goal is to have the LLM accurately identify variables from the simulation selected by a teacher that best matches their IGs. In this paper, we empirically evaluate this task's accuracy on 63 physics simulations from PhET and Golabz using two LLMs: ChatGPT-3.5 \cite{gpt3} and PaLM 2 \cite{anil2023palm2}. By employing the recently introduced TELeR taxonomy, we examined the impact of different prompting strategies on LLM's ability to identify the physical variables relevant to the IGs. Our findings demonstrated that SimPal can perform this task with a high degree of accuracy when provided with an appropriately crafted prompt.      

\section{Background and Related Work}

\textbf{Conversational Agents in K-12 Science.}
Conversational agents, like  Betty's Brain \citep{kinnebrew2012identifying,kinnebrew2013contextualized} and MetaTutor \citep{Metatutor, metatutorDef} have been used to foster students' learning. In Betty's Brain \citep{kinnebrew2012identifying,kinnebrew2013contextualized}, students learn science and mathematics concepts by teaching a virtual agent, Betty. 
\textit{MetaTutor} is a hypermedia-based biology learning environment where teachers set learning goals and students choose metacognitive processes, with occasional pedagogical agent prompts. All of the aforementioned frameworks support students' learning, whereas SimPal offers a conversational AI assistant for teachers to develop simulation-based science lesson plans.

\smallskip
\noindent \textbf{LLMs and K-12 Education.}
LLMs have recently been increasingly used to enhance student learning. Zhang et al. utilized LLMs in solving arithmetic math word problems \cite{zhang2023ACL}.
Prihar et al. ~\cite{prihar2023comparing} utilized GPT-3 with few shot learning to generate middle school math explanations on ASSISTments. They found that GPT-3, primarily trained on English text, generated explanations that were significantly inferior to teacher-authored ones. Lately, Khan Academy has introduced a GPT-4 \cite{gpt4} powered tutoring system, Khanmigo \cite{Khanmigo}, to assist teachers in planning their lessons and providing feedback on students writing.
Our proposed approach, SimPal, is similar to Khanmigo in terms of assisting teachers in planning their lessons.
However,  SimPal differs from Khanmigo in that it allows teachers to integrate any \textit{third-party simulations} into their lesson plans.

\smallskip
\noindent \textbf{Grounding LLMs to Unseen Tasks.} LLMs, which represent vast amounts of information, still require adaptation to specific tasks. Traditionally, task-specific supervised data is used to fine-tune an LLM and adapt it to new natural language processing (NLP) applications~\cite{dai2015semi, howard2018universal,radford2019language, hu2023meta}. However, fine-tuning faces two major challenges: insufficient training data and a lack of computing resources and expertise. Few-shot learning is another approach that uses prompt engineering~\cite{gao2021making,chen2022decoupling} and domain-specific examples~\cite{gpt3}. However, few-shot learning may be challenging for lesson planning due to teachers' individual teaching styles and preferences. Reinforcement learning (RL) from human feedback (RLHF) employs RL to optimize human preferences during LLM training \cite{ouyang2022training}. However, it can incur significant exploration costs in RL. In contrast, our approach, known as \textit{meta-conversation}, uses natural conversation to infer a human preference, i.e., the teacher’s lesson plan.

\smallskip
\noindent \textbf{Prompt Taxonomy for LLM.} As LLM's prompt impacts the output accuracy of LLMs, a recent study proposed a taxonomy, TELeR \cite{santu2023teler}, to design and evaluate prompting techniques systematically. TELeR taxonomy has seven levels of prompts. We only explain the four prompt levels [Level 1- Level 4] used in our study in Table~\ref{TeLER}. 

\section{Instruction Goals and SimPal}
We formulate a teacher's IG in terms of variables and relationships among variables. 
Consider a toy example where the teacher's instructional goal is to teach inversely proportional relationships in Newton's Second Law of Motion in a PhET simulation \cite{Newton2ndLaw}. As demonstrated in Figure~\ref{fig:Intro}, the teacher conveys their IGs (e.g., inversely proportional relationships Newton's Second Law of Motion) to SimPal. Then, SimPal generates relevant topics (e.g., force, acceleration) for the lab and asks the teacher to review those. Upon receiving the teacher's feedback, SimPal then identifies a set of relevant variables and their relationships to create symbolic representations of the desired IGs based on the teacher's feedback.

The scope of our study is variable extraction in Physics simulations, with the task described as follows.

\noindent \textbf{Problem Definition.}
Given an IG of a simulation topic, SimPal uses LLMs to generate \textit{variables}. The task is to assess LLM’s accuracy of generated variables given a natural language description of the IG. 

\begin{table}
\centering

\caption{TELeR Taxonomy for LLM Prompting}\label{TeLER}
{
\begin{tabular}{l|l}
\hline
Level (L) &  Definition \\
\hline\hline
L1 & One sentence describing the high-level task goal\\\hline
L2 & Multi-sentence prompt describing the high-level   \\
 &  goals and sub-tasks \\\hline

L3 & Prompt describing the high-level goals  \\
 &  and sub-tasks in bulleted style.\\\hline
L4 & Prompt specifying high-level goals, sub-tasks, and   \\
 & output evaluation criteria (e.g., few-shot examples)  \\\hline

\hline

\end{tabular}
}
\end{table}

\section{Experimental Design}

\subsection{Underlying LLM of SimPal}
Table~\ref{LLMEvaluated} lists three LLMs that we assessed in our preliminary analysis.

\begin{table}[!htb]
\caption{LLMs Evaluated in this work.}\label{LLMEvaluated}
\begin{tabular}{c| c| c}
\hline
Model &  Creator & \# Parameters  \\\hline\hline
ChatGPT-3.5 (\texttt{gpt-3.5-turbo-0613}, \cite{gpt3})& OpenAI & 175B     \\
PaLM 2 (\texttt{chat-bison-001}, \cite{anil2023palm2})& Google & 340B  \\
 LLaMA-2 (\texttt{Llama-2-70b-chat-hf}, \cite{llama})  & Meta & 70B  \\\hline
\end{tabular}
\end{table}

\subsection{Prompt Design with SimPal} 

We used Level 1 to Level 4 following the TELeR taxonomy in Table \ref{TeLER}. Example Level 1, 2, 3, and 4 prompts are given below. 
\begin{itemize}[leftmargin=*]
\item \textbf{Level 1} Identify and list the variables associated with these topics and the description, along with their corresponding symbols.

\item \textbf{Level 2} You are a physics teacher in a high school, and you are preparing a lesson plan on related concepts. You have a list of topics and descriptions. \\
Your task is to \textbf{\textit{Level 1 Prompt Text} }\\
Please provide the variables and symbols in the following JSON format. The key would be the ``Name" of the variable and the value would be the ``Symbol". \\
Include symbols and strictly follow the JSON format. \\
Do not print topics and descriptions; only variable names and corresponding symbols are used.
\item \textbf{Level 3} \textbf{\textit{Level 2 Prompt Text}}\\
Please provide the variables and symbols in the following JSON format:
 [
    { ``Name":  " ", ``Symbol": " "}
]\\
- List down all the relevant variables and their symbols. 

\item \textbf{Level 4} \textbf{\textit{Level 3 Prompt Text}}\\
 You are given a GUIDELINES\_PROMPT to show an example but do not include the variables from the GUIDELINES\_PROMPT in the response if they are not relevant. 
\end{itemize}


\subsection{Simulation Dataset} 
Our dataset includes simulations from PhET \cite{PhET} and Golabz \cite{Golabz}. PhET hosts free math and science simulations. Golabz hosts online science labs to promote inquiry learning at scale. We performed preliminary analysis on five PhET simulations (Section \ref{PrelimExperiment}) and final evaluation on 32 PhET and 31 Golabz simulations (Section \ref{FinalEval}).

\subsection{Preliminary Experiments and Insights}\label{PrelimExperiment}
We investigated the output of three LLMs on five PhET simulations using the TELeR taxonomy prompting levels [Level 1– Level 4]. Table \ref{Levels} shows that all three LLMs' F1-scores fall with Level-4 prompting. Observing the format accuracy of Levels 2 and 3, we
conclude that ChatGPT-3.5 and PaLM 2 generate output in the desired format. Based on the results in Table \ref{Levels}, we selected two LLMs, ChatGPT-3.5 and PaLM 2, with Level 2 and Level 3 prompting levels.

\begin{table}[!htb]
\centering 
\caption{LLM Performance and Prompting Levels as per the TeLER Taxonomy. Format Accuracy =  (0) 1, if LLM-generated Results (Do not) Follow the Prompt's Format Specification. The Highest of each Metric per Prompt Level is in Bold}
\label{Levels}
\begin{tabular}{c| c| c| c |c}
\hline
Model &  Format Accuracy & Precision & Recall & F1 Score \\\hline\hline
\multicolumn{2}{c}{Level 1}  \\\hline
ChatGPT-3.5 &  0 & 0.923 & 0.923 & 0.923 \\
PaLM 2 & 0 & 0.923 & 0.958 & 0.94 \\
 LLaMA-2 (70B) & 0 & \textbf{0.929} & \textbf{1} & \textbf{0.963} \\\hline
 \multicolumn{2}{c}{Level 2}  \\\hline
 ChatGPT-3.5 & 1 & 0.78 & 0.729 & 0.754\\
PaLM 2 & 1 & \textbf{0.881} & 0.835 & 0.857\\
 LLaMA-2 (70B) & 0 & 0.876 & \textbf{0.897} & \textbf{0.887} \\\hline
 \multicolumn{2}{c}{Level 3}  \\\hline
 ChatGPT-3.5 & 1 & \textbf{0.898} & \textbf{0.877} & \textbf{0.887} \\
PaLM 2 & 1 & 0.853 & 0.848 & 0.851 \\
 LLaMA-2 (70B) & 0.4 & 0.755 & 0.767 & 0.761 \\\hline
 \multicolumn{2}{c}{Level 4}  \\\hline
  ChatGPT-3.5 & 1 & 0.732 & 0.691 & 0.711 \\
PaLM 2& 1 & \textbf{0.96} & 0.712 & \textbf{0.818}\\
 LLaMA-2 (70B) & 0 & 0.82 & \textbf{0.761} & 0.7894\\
\hline
\end{tabular}

\end{table}

\section{Final Case Study and Evaluation}\label{FinalEval}

\textbf{Dataset.}  
We evaluated  SimPal's performance
 in 63 Physics simulations, including 32 from PhET and 31 from Golabz, as depicted in Table~\ref{Dataset}.
For each simulation, we designed two prompting levels (Level 2 and Level 3) using two LLMs: ChatGPT-3.5 and PaLM 2.

\begin{table}[!htb]
\centering
\caption{Dataset Statistics. L2 = Level 2, L3 = Level 3, \#Prompts = Total Prompts by Level 2 and Level 3}
\label{Dataset}
\begin{tabular}{c| c| c| c |c|c|c}
\hline
&\multicolumn{3}{|c|}{ChatGPT-3.5} &  \multicolumn{3}{c}{PaLM 2}  \\\hline
 &  L2 & L3 &\#Prompts  & L2 & L3 &  \#Prompts  \\\hline\hline

Golabz &  32 & 32 & 64 & 32 & 32 &64\\
PhET &31 & 31& 62 &31 & 31& 62\\\hline

\end{tabular}

\end{table}

\noindent \textbf{Evaluation.} We created prompts by extracting IGs and topics from lab web pages. The IGs in PhET and Golabz are the learning goals and lab descriptions, respectively. To identify gold standard
variables for a lab, we identified topics from the lab webpage and added additional terms from the Teacher Resources section. Finally, we cross-referenced the relevant terms with an open-source CK-12 Physical Science textbook \cite {CKPhysics}, aligned to the Next Generation Science Standards (NGSS) \cite{national2013next} to determine the final gold standards and manually compared SimPal's outputs to the gold standards.

\noindent \textbf{Metric.} For each simulation, the LLM inferred variables are com-
pared against the list of gold standard variables to compute the true
positive, false positive, true negative, and false negative statistics. Then, all such statistics in a dataset
were aggregated  to compute the final Precision, Recall,
and micro-averaged F1 score.

\begin{table}[h!]
\caption{An Example Annotation Scheme and SimPal's Output Evaluation on a Lab Titled \textit{Wave on a String}}\label{Annotation}
{
\begin{tabular}{l|l|l}
\hline
 Topics & LLM Output & Gold Standard\\
\hline\hline

   & { ""Name"": ""Wavelength""},"&  frequency\\
  &{, ""Symbol"": ""\u03BB"" }& \\
 
  Frequency & { ""Name"": ""Frequency""},& amplitude\\
 
  & { ""Symbol"": ""f"" }& \\
   Amplitude & { ""Name"": ""Period"" },  & wavelength\\

   & { ""Symbol"": ""T"" }& \\
  
    Damping &  { ""Name"": ""Amplitude"" },& period\\
    & { ""Symbol"": ""A"" }& \\
         & { ""Name"": ""Speed"" }, & \\
       & { ""Symbol"": ""v"" }& \\
           & { ""Name"": ""Damping Coefficient"",} & \\
             & { ""Symbol"": ""\u03B2"" } & \\\hline
\end{tabular}

}
\end{table}

Table \ref{Annotation} presents an example of SimPal's output evaluation in a lab. We calculated true positive values (TP) by comparing the number of matched LLM outputs to the gold standard, resulting in four true positives. We calculated false positives (FP) by subtracting the number of LLM outputs from the true positives, yielding two false positives. Further, we calculated the false negatives (FN) by subtracting true positives from the number of gold standard outputs, resulting in zero false negatives in the given example.
\subsection{Results and Discussion}
Table \ref{Results} presents our evaluation results of SimPal. 

\smallskip
\noindent \textbf{TELeR Prompting Levels and SimPal Performance.}
Level 3 prompting resulted in higher F1 scores for both LLMs than Level 2 in Golabz simulations. In PhET simulations, Level 2 prompting produced a higher recall score than Level 3 in PaLM 2.

\smallskip
\noindent \textbf{LLM Family and SimPal Performance.} ChatGPT-3.5 outperformed PaLM 2 in F1-scores in both Golabz and PhET simulations with Level 3 prompting. ChatGPT-3.5 also achieved a higher F1 score than PaLM 2 for Level 2 prompting in Golabz simulations.

\smallskip
\noindent \textbf{Simulation Source and SimPal Performance.} Golabz simulations resulted in a higher F1-score in both Level 2 and Level 3 prompting than PhET in 
ChatGPT-3.5. In PaLM 2, Golabz simulations outperformed PhET in F1-score in only Level 3 prompting. 

\smallskip
The differences in F1 scores between Golabz and PhET simulations may be due to content alignment differences. Golabz simulations may have been more aligned with curriculum standards. Additionally, PhET simulations may contain more complex or detailed information, resulting in the generation of extraneous outputs.
\begin{table}
\centering
\caption{SimPal's Performance with TeLER Prompt Levels 2 and 3 for LLM Families and Simulation Sources in Table \ref{Dataset} }\label{Results}
\begin{tabular}{c| c|c|c|c|c|c}
\hline
\multicolumn{7}{c}{ChatGPT-3.5}  \\\hline\hline
 &  Precision & Recall & F1  & Precision & Recall & F1  \\\hline
&\multicolumn{3}{c|}{Level 3} &  \multicolumn{3}{c}{Level 2}  \\\hline
Golabz &  0.590 & 0.713 & 0.60 & 0.525 & 0.627 &0.541\\
PhET &0.560 & 0.654 & 0.581 & 0.523 & 0.519 & 0.539\\\hline
\multicolumn{7}{c}{PaLM 2}  \\\hline\hline
&\multicolumn{3}{c|}{Level 3} &  \multicolumn{3}{c}{Level 2}  \\\hline
 Golabz& 0.607 & 0.639 & 0.568 & 0.555 & 0.591 & 0.525\\
 PhET & 0.512 & 0.584 & 0.547 & 0.529 & 0.628 &0.547\\\hline
\end{tabular}
\end{table}

\section{Future Work}
We plan to extend SimPal to provide support to students via meta-conversation. This includes feedback on writings, answered questions, and hint generation. Additionally, we plan to use SimPal's student interaction data to generate recommendations for teachers, such as identifying high-performing and struggling students.

\section{Conclusion}
In this study, we present SimPal, an LLM-based meta-conversational framework for simulation-based science labs, allowing teachers to include third-party (open or closed-source) simulations into lesson plans, facilitating \textit{instruction at scale.} We assessed SimPal's variable generation capabilities with two LLMs: ChatGPT-3.5 and PaLM 2 on 63 Physics simulations from PhET and Golabz, experimenting with different prompts following the TELeR prompting taxonomy. Our findings showed that I) SimPal can provide a meaningful variable list tailored to the lab and instruction goal, and II) the LLM prompting level impacts SimPal's performance. Furthermore, we observed that Golabz simulations outperformed PhET in the F1 score. It is important to note a limitation in our evaluation; our gold standard outputs may lack the subject matter expertise of real school teachers, potentially leading to disparities in F1 scores. Future work will involve incorporating feedback from teachers and subject matter experts to improve the accuracy and relevance of LLM outputs.

\begin{acks}
This work was funded in part by the National Science Foundation through Grant 2302974.
\end{acks}

 \bibliographystyle{ACM-Reference-Format}
 \balance
 \bibliography{SimPal}


\begin{thebibliography}{34}


\ifx \showCODEN    \undefined \def \showCODEN     #1{\unskip}     \fi
\ifx \showDOI      \undefined \def \showDOI       #1{#1}\fi
\ifx \showISBNx    \undefined \def \showISBNx     #1{\unskip}     \fi
\ifx \showISBNxiii \undefined \def \showISBNxiii  #1{\unskip}     \fi
\ifx \showISSN     \undefined \def \showISSN      #1{\unskip}     \fi
\ifx \showLCCN     \undefined \def \showLCCN      #1{\unskip}     \fi
\ifx \shownote     \undefined \def \shownote      #1{#1}          \fi
\ifx \showarticletitle \undefined \def \showarticletitle #1{#1}   \fi
\ifx \showURL      \undefined \def \showURL       {\relax}        \fi
\providecommand\bibfield[2]{#2}
\providecommand\bibinfo[2]{#2}
\providecommand\natexlab[1]{#1}
\providecommand\showeprint[2][]{arXiv:#2}

\bibitem[Achiam et~al\mbox{.}(2023)]%
        {gpt4}
\bibfield{author}{\bibinfo{person}{Josh Achiam}, \bibinfo{person}{Steven Adler}, \bibinfo{person}{Sandhini Agarwal}, \bibinfo{person}{Lama Ahmad}, \bibinfo{person}{Ilge Akkaya}, \bibinfo{person}{Florencia~Leoni Aleman}, \bibinfo{person}{Diogo Almeida}, \bibinfo{person}{Janko Altenschmidt}, \bibinfo{person}{Sam Altman}, \bibinfo{person}{Shyamal Anadkat}, {et~al\mbox{.}}} \bibinfo{year}{2023}\natexlab{}.
\newblock \showarticletitle{Gpt-4 technical report}.
\newblock \bibinfo{journal}{\emph{arXiv preprint arXiv:2303.08774}} (\bibinfo{year}{2023}).
\newblock


\bibitem[Anil et~al\mbox{.}(2023)]%
        {anil2023palm2}
\bibfield{author}{\bibinfo{person}{Rohan Anil}, \bibinfo{person}{Andrew~M Dai}, \bibinfo{person}{Orhan Firat}, \bibinfo{person}{Melvin Johnson}, \bibinfo{person}{Dmitry Lepikhin}, \bibinfo{person}{Alexandre Passos}, \bibinfo{person}{Siamak Shakeri}, \bibinfo{person}{Emanuel Taropa}, \bibinfo{person}{Paige Bailey}, \bibinfo{person}{Zhifeng Chen}, {et~al\mbox{.}}} \bibinfo{year}{2023}\natexlab{}.
\newblock \showarticletitle{Palm 2 technical report}.
\newblock \bibinfo{journal}{\emph{arXiv preprint arXiv:2305.10403}} (\bibinfo{year}{2023}).
\newblock


\bibitem[Azevedo et~al\mbox{.}(2009)]%
        {metatutorDef}
\bibfield{author}{\bibinfo{person}{Roger Azevedo}, \bibinfo{person}{Amy~M Witherspoon}, \bibinfo{person}{Arthur~C Graesser}, \bibinfo{person}{Danielle~S McNamara}, \bibinfo{person}{Amber Chauncey}, {and} \bibinfo{person}{Emily Siler}.} \bibinfo{year}{2009}\natexlab{}.
\newblock \showarticletitle{MetaTutor: Analyzing Self-Regulated Learning in a Tutoring System for Biology.} \bibinfo{publisher}{AIED}.
\newblock


\bibitem[Bouchet et~al\mbox{.}(2012)]%
        {Metatutor}
\bibfield{author}{\bibinfo{person}{Fran{\c{c}}ois Bouchet}, \bibinfo{person}{Roger Azevedo}, \bibinfo{person}{John~S Kinnebrew}, {and} \bibinfo{person}{Gautam Biswas}.} \bibinfo{year}{2012}\natexlab{}.
\newblock \showarticletitle{Identifying Students' Characteristic Learning Behaviors in an Intelligent Tutoring System Fostering Self-Regulated Learning.}
\newblock \bibinfo{journal}{\emph{International Educational Data Mining Society}} (\bibinfo{year}{2012}).
\newblock


\bibitem[Brown et~al\mbox{.}(2020)]%
        {gpt3}
\bibfield{author}{\bibinfo{person}{Tom Brown}, \bibinfo{person}{Benjamin Mann}, \bibinfo{person}{Nick Ryder}, \bibinfo{person}{Melanie Subbiah}, \bibinfo{person}{Jared~D Kaplan}, \bibinfo{person}{Prafulla Dhariwal}, \bibinfo{person}{Arvind Neelakantan}, \bibinfo{person}{Pranav Shyam}, \bibinfo{person}{Girish Sastry}, \bibinfo{person}{Amanda Askell}, {et~al\mbox{.}}} \bibinfo{year}{2020}\natexlab{}.
\newblock \showarticletitle{Language models are few-shot learners}.
\newblock \bibinfo{journal}{\emph{Advances in neural information processing systems}}  \bibinfo{volume}{33} (\bibinfo{year}{2020}), \bibinfo{pages}{1877--1901}.
\newblock


\bibitem[Chen et~al\mbox{.}(2022)]%
        {chen2022decoupling}
\bibfield{author}{\bibinfo{person}{Xiang Chen}, \bibinfo{person}{Lei Li}, \bibinfo{person}{Ningyu Zhang}, \bibinfo{person}{Xiaozhuan Liang}, \bibinfo{person}{Shumin Deng}, \bibinfo{person}{Chuanqi Tan}, \bibinfo{person}{Fei Huang}, \bibinfo{person}{Luo Si}, {and} \bibinfo{person}{Huajun Chen}.} \bibinfo{year}{2022}\natexlab{}.
\newblock \showarticletitle{Decoupling knowledge from memorization: Retrieval-augmented prompt learning}.
\newblock \bibinfo{journal}{\emph{Advances in Neural Information Processing Systems}}  \bibinfo{volume}{35} (\bibinfo{year}{2022}), \bibinfo{pages}{23908--23922}.
\newblock


\bibitem[Chowdhery et~al\mbox{.}(2023)]%
        {chowdhery2023palm}
\bibfield{author}{\bibinfo{person}{Aakanksha Chowdhery}, \bibinfo{person}{Sharan Narang}, \bibinfo{person}{Jacob Devlin}, \bibinfo{person}{Maarten Bosma}, \bibinfo{person}{Gaurav Mishra}, \bibinfo{person}{Adam Roberts}, \bibinfo{person}{Paul Barham}, \bibinfo{person}{Hyung~Won Chung}, \bibinfo{person}{Charles Sutton}, \bibinfo{person}{Sebastian Gehrmann}, {et~al\mbox{.}}} \bibinfo{year}{2023}\natexlab{}.
\newblock \showarticletitle{Palm: Scaling language modeling with pathways}.
\newblock \bibinfo{journal}{\emph{Journal of Machine Learning Research}} \bibinfo{volume}{24}, \bibinfo{number}{240} (\bibinfo{year}{2023}), \bibinfo{pages}{1--113}.
\newblock


\bibitem[CK-12(2024)]%
        {CKPhysics}
\bibfield{author}{\bibinfo{person}{CK-12}.} \bibinfo{year}{2024}\natexlab{}.
\newblock \bibinfo{title}{CK-12 Physical Science for Middle School}.
\newblock \bibinfo{howpublished}{https://flexbooks.ck12.org/cbook/ck-12-middle-school-physical-science-flexbook-2.0/}.
\newblock


\bibitem[Council et~al\mbox{.}(2013)]%
        {national2013next}
\bibfield{author}{\bibinfo{person}{National~Research Council} {et~al\mbox{.}}} \bibinfo{year}{2013}\natexlab{}.
\newblock \showarticletitle{Next generation science standards: For states, by states}.
\newblock  (\bibinfo{year}{2013}).
\newblock


\bibitem[Dai and Le(2015)]%
        {dai2015semi}
\bibfield{author}{\bibinfo{person}{Andrew~M Dai} {and} \bibinfo{person}{Quoc~V Le}.} \bibinfo{year}{2015}\natexlab{}.
\newblock \showarticletitle{Semi-supervised sequence learning}.
\newblock \bibinfo{journal}{\emph{Advances in neural information processing systems}}  \bibinfo{volume}{28} (\bibinfo{year}{2015}).
\newblock


\bibitem[Falloon(2019)]%
        {StudentStruggleSimulation}
\bibfield{author}{\bibinfo{person}{Garry Falloon}.} \bibinfo{year}{2019}\natexlab{}.
\newblock \showarticletitle{Using simulations to teach young students science concepts: An Experiential Learning theoretical analysis}.
\newblock \bibinfo{journal}{\emph{Computers \& Education}}  \bibinfo{volume}{135} (\bibinfo{year}{2019}), \bibinfo{pages}{138--159}.
\newblock


\bibitem[Fischer and Dershimer(2020)]%
        {TeacherSimulationStruggle}
\bibfield{author}{\bibinfo{person}{Christian Fischer} {and} \bibinfo{person}{R~Charles Dershimer}.} \bibinfo{year}{2020}\natexlab{}.
\newblock \showarticletitle{Preparing teachers to use educational games, virtual experiments, and interactive science simulations for engaging students in the practices of science}.
\newblock  (\bibinfo{year}{2020}).
\newblock


\bibitem[Gao et~al\mbox{.}(2021)]%
        {gao2021making}
\bibfield{author}{\bibinfo{person}{Tianyu Gao}, \bibinfo{person}{Adam Fisch}, {and} \bibinfo{person}{Danqi Chen}.} \bibinfo{year}{2021}\natexlab{}.
\newblock \showarticletitle{Making Pre-trained Language Models Better Few-shot Learners}. In \bibinfo{booktitle}{\emph{Proceedings of the 59th Annual Meeting of the Association for Computational Linguistics and the 11th International Joint Conference on Natural Language Processing (Volume 1: Long Papers)}}. \bibinfo{pages}{3816--3830}.
\newblock


\bibitem[Gonz{\'a}lez-Cruz et~al\mbox{.}(2003)]%
        {gonzalez2003convenience}
\bibfield{author}{\bibinfo{person}{Javier Gonz{\'a}lez-Cruz}, \bibinfo{person}{Rogelio Rodr{\'\i}guez-Sotres}, {and} \bibinfo{person}{Mireya Rodr{\'\i}guez-Penagos}.} \bibinfo{year}{2003}\natexlab{}.
\newblock \showarticletitle{On the convenience of using a computer simulation to teach enzyme kinetics to undergraduate students with biological chemistry-related curricula}.
\newblock \bibinfo{journal}{\emph{Biochemistry and Molecular Biology Education}} \bibinfo{volume}{31}, \bibinfo{number}{2} (\bibinfo{year}{2003}), \bibinfo{pages}{93--101}.
\newblock


\bibitem[Graesser et~al\mbox{.}(2006)]%
        {StudentStruggleSimulation2}
\bibfield{author}{\bibinfo{person}{Arthur~C Graesser}, \bibinfo{person}{G~Tanner Jackson}, \bibinfo{person}{Hyun-Jeong~Joyce Kim}, {and} \bibinfo{person}{Andrew Olney}.} \bibinfo{year}{2006}\natexlab{}.
\newblock \showarticletitle{AutoTutor 3-D Simulations: Analyzing Users' Actions and Learning Trends}. In \bibinfo{booktitle}{\emph{Proceedings of the Annual Meeting of the Cognitive Science Society}}, Vol.~\bibinfo{volume}{28}.
\newblock


\bibitem[Howard and Ruder(2018)]%
        {howard2018universal}
\bibfield{author}{\bibinfo{person}{Jeremy Howard} {and} \bibinfo{person}{Sebastian Ruder}.} \bibinfo{year}{2018}\natexlab{}.
\newblock \showarticletitle{Universal Language Model Fine-tuning for Text Classification}. In \bibinfo{booktitle}{\emph{Proceedings of the 56th Annual Meeting of the Association for Computational Linguistics (Volume 1: Long Papers)}}. \bibinfo{pages}{328--339}.
\newblock


\bibitem[Hu et~al\mbox{.}(2023)]%
        {hu2023meta}
\bibfield{author}{\bibinfo{person}{Nathan Hu}, \bibinfo{person}{Eric Mitchell}, \bibinfo{person}{Christopher~D Manning}, {and} \bibinfo{person}{Chelsea Finn}.} \bibinfo{year}{2023}\natexlab{}.
\newblock \showarticletitle{Meta-Learning Online Adaptation of Language Models}. In \bibinfo{booktitle}{\emph{Proceedings of the 2023 Conference on Empirical Methods in Natural Language Processing}}. \bibinfo{pages}{4418--4432}.
\newblock


\bibitem[Khan(2024)]%
        {Khanmigo}
\bibfield{author}{\bibinfo{person}{Sal Khan}.} \bibinfo{year}{2024}\natexlab{}.
\newblock \bibinfo{title}{Khanmigo.}
\newblock \bibinfo{howpublished}{https://www.khanacademy.org/khan-labs}.
\newblock


\bibitem[Kinnebrew and Biswas(2012)]%
        {kinnebrew2012identifying}
\bibfield{author}{\bibinfo{person}{John~S Kinnebrew} {and} \bibinfo{person}{Gautam Biswas}.} \bibinfo{year}{2012}\natexlab{}.
\newblock \showarticletitle{Identifying Learning Behaviors by Contextualizing Differential Sequence Mining with Action Features and Performance Evolution.}
\newblock \bibinfo{journal}{\emph{International Educational Data Mining Society}} (\bibinfo{year}{2012}).
\newblock


\bibitem[Kinnebrew et~al\mbox{.}(2013)]%
        {kinnebrew2013contextualized}
\bibfield{author}{\bibinfo{person}{John~S Kinnebrew}, \bibinfo{person}{Kirk~M Loretz}, {and} \bibinfo{person}{Gautam Biswas}.} \bibinfo{year}{2013}\natexlab{}.
\newblock \showarticletitle{A contextualized, differential sequence mining method to derive students' learning behavior patterns}.
\newblock \bibinfo{journal}{\emph{JEDM| Journal of Educational Data Mining}} \bibinfo{volume}{5}, \bibinfo{number}{1} (\bibinfo{year}{2013}), \bibinfo{pages}{190--219}.
\newblock


\bibitem[Koll{\"o}ffel and De~Jong(2013)]%
        {simulations2}
\bibfield{author}{\bibinfo{person}{Bas Koll{\"o}ffel} {and} \bibinfo{person}{Ton De~Jong}.} \bibinfo{year}{2013}\natexlab{}.
\newblock \showarticletitle{Conceptual Understanding of electrical circuits in secondary vocational engineering education: Combining traditional instruction with inquiry learning in a virtual lab}.
\newblock \bibinfo{journal}{\emph{Journal of engineering education}} \bibinfo{volume}{102}, \bibinfo{number}{3} (\bibinfo{year}{2013}), \bibinfo{pages}{375--393}.
\newblock


\bibitem[Lecerio(2019)]%
        {Newton2ndLaw}
\bibfield{author}{\bibinfo{person}{Leah~L. Lecerio}.} \bibinfo{year}{2019}\natexlab{}.
\newblock \bibinfo{title}{Newton's Second Law of Motion}.
\newblock \bibinfo{howpublished}{https://phet.colorado.edu/en/contributions/view/5092}.
\newblock


\bibitem[Li et~al\mbox{.}(2021)]%
        {LLMQgen}
\bibfield{author}{\bibinfo{person}{Alexander~Hanbo Li}, \bibinfo{person}{Patrick Ng}, \bibinfo{person}{Peng Xu}, \bibinfo{person}{Henghui Zhu}, \bibinfo{person}{Zhiguo Wang}, {and} \bibinfo{person}{Bing Xiang}.} \bibinfo{year}{2021}\natexlab{}.
\newblock \showarticletitle{Dual Reader-Parser on Hybrid Textual and Tabular Evidence for Open Domain Question Answering}. In \bibinfo{booktitle}{\emph{Proceedings of the 59th Annual Meeting of the Association for Computational Linguistics and the 11th International Joint Conference on Natural Language Processing (Volume 1: Long Papers)}}. \bibinfo{pages}{4078--4088}.
\newblock


\bibitem[Ouyang et~al\mbox{.}(2022)]%
        {ouyang2022training}
\bibfield{author}{\bibinfo{person}{Long Ouyang}, \bibinfo{person}{Jeffrey Wu}, \bibinfo{person}{Xu Jiang}, \bibinfo{person}{Diogo Almeida}, \bibinfo{person}{Carroll Wainwright}, \bibinfo{person}{Pamela Mishkin}, \bibinfo{person}{Chong Zhang}, \bibinfo{person}{Sandhini Agarwal}, \bibinfo{person}{Katarina Slama}, \bibinfo{person}{Alex Ray}, {et~al\mbox{.}}} \bibinfo{year}{2022}\natexlab{}.
\newblock \showarticletitle{Training language models to follow instructions with human feedback}.
\newblock \bibinfo{journal}{\emph{Advances in neural information processing systems}}  \bibinfo{volume}{35} (\bibinfo{year}{2022}), \bibinfo{pages}{27730--27744}.
\newblock


\bibitem[Park et~al\mbox{.}(2023)]%
        {park2023integrating}
\bibfield{author}{\bibinfo{person}{Joonhyeong Park}, \bibinfo{person}{Tang~Wee Teo}, \bibinfo{person}{Arnold Teo}, \bibinfo{person}{Jina Chang}, \bibinfo{person}{Jun~Song Huang}, {and} \bibinfo{person}{Sengmeng Koo}.} \bibinfo{year}{2023}\natexlab{}.
\newblock \showarticletitle{Integrating artificial intelligence into science lessons: teachers’ experiences and views}.
\newblock \bibinfo{journal}{\emph{International Journal of STEM Education}} \bibinfo{volume}{10}, \bibinfo{number}{1} (\bibinfo{year}{2023}), \bibinfo{pages}{61}.
\newblock


\bibitem[Prihar et~al\mbox{.}(2023)]%
        {prihar2023comparing}
\bibfield{author}{\bibinfo{person}{Ethan Prihar}, \bibinfo{person}{Morgan Lee}, \bibinfo{person}{Mia Hopman}, \bibinfo{person}{Adam~Tauman Kalai}, \bibinfo{person}{Sofia Vempala}, \bibinfo{person}{Allison Wang}, \bibinfo{person}{Gabriel Wickline}, \bibinfo{person}{Aly Murray}, {and} \bibinfo{person}{Neil Heffernan}.} \bibinfo{year}{2023}\natexlab{}.
\newblock \showarticletitle{Comparing different approaches to generating mathematics explanations using large language models}. In \bibinfo{booktitle}{\emph{International Conference on Artificial Intelligence in Education}}. Springer, \bibinfo{pages}{290--295}.
\newblock


\bibitem[Radford et~al\mbox{.}(2019)]%
        {radford2019language}
\bibfield{author}{\bibinfo{person}{Alec Radford}, \bibinfo{person}{Jeffrey Wu}, \bibinfo{person}{Rewon Child}, \bibinfo{person}{David Luan}, \bibinfo{person}{Dario Amodei}, \bibinfo{person}{Ilya Sutskever}, {et~al\mbox{.}}} \bibinfo{year}{2019}\natexlab{}.
\newblock \showarticletitle{Language models are unsupervised multitask learners}.
\newblock \bibinfo{journal}{\emph{OpenAI blog}} \bibinfo{volume}{1}, \bibinfo{number}{8} (\bibinfo{year}{2019}), \bibinfo{pages}{9}.
\newblock


\bibitem[Rutten et~al\mbox{.}(2012)]%
        {Simulations}
\bibfield{author}{\bibinfo{person}{Nico Rutten}, \bibinfo{person}{Wouter~R Van~Joolingen}, {and} \bibinfo{person}{Jan~T Van Der~Veen}.} \bibinfo{year}{2012}\natexlab{}.
\newblock \showarticletitle{The learning effects of computer simulations in science education}.
\newblock \bibinfo{journal}{\emph{Computers \& education}} \bibinfo{volume}{58}, \bibinfo{number}{1} (\bibinfo{year}{2012}), \bibinfo{pages}{136--153}.
\newblock


\bibitem[Santu and Feng(2023)]%
        {santu2023teler}
\bibfield{author}{\bibinfo{person}{Shubhra Kanti~Karmaker Santu} {and} \bibinfo{person}{Dongji Feng}.} \bibinfo{year}{2023}\natexlab{}.
\newblock \showarticletitle{TELeR: A General Taxonomy of LLM Prompts for Benchmarking Complex Tasks}. In \bibinfo{booktitle}{\emph{Findings of the Association for Computational Linguistics: EMNLP}}. \bibinfo{pages}{14197–--14203}.
\newblock


\bibitem[Sundar and Heck(2023)]%
        {LLMConversation}
\bibfield{author}{\bibinfo{person}{Anirudh~S Sundar} {and} \bibinfo{person}{Larry Heck}.} \bibinfo{year}{2023}\natexlab{}.
\newblock \showarticletitle{cTBLS: Augmenting Large Language Models with Conversational Tables}. In \bibinfo{booktitle}{\emph{Proceedings of the 5th Workshop on NLP for Conversational AI (NLP4ConvAI 2023)}}. \bibinfo{pages}{59--70}.
\newblock


\bibitem[Touvron et~al\mbox{.}(2023)]%
        {llama}
\bibfield{author}{\bibinfo{person}{Hugo Touvron}, \bibinfo{person}{Louis Martin}, \bibinfo{person}{Kevin Stone}, \bibinfo{person}{Peter Albert}, \bibinfo{person}{Amjad Almahairi}, \bibinfo{person}{Yasmine Babaei}, \bibinfo{person}{Nikolay Bashlykov}, \bibinfo{person}{Soumya Batra}, \bibinfo{person}{Prajjwal Bhargava}, \bibinfo{person}{Shruti Bhosale}, {et~al\mbox{.}}} \bibinfo{year}{2023}\natexlab{}.
\newblock \showarticletitle{Llama 2: Open foundation and fine-tuned chat models}.
\newblock \bibinfo{journal}{\emph{arXiv preprint arXiv:2307.09288}} (\bibinfo{year}{2023}).
\newblock


\bibitem[University~of Twente(2012)]%
        {Golabz}
\bibfield{author}{\bibinfo{person}{the~Netherlands University~of Twente}.} \bibinfo{year}{2012}\natexlab{}.
\newblock \bibinfo{title}{Global Online Science Labs for Inquiry Learning in Schools}.
\newblock \bibinfo{howpublished}{https://premium.golabz.eu/about/go-lab-initiative}.
\newblock


\bibitem[Wieman(2002)]%
        {PhET}
\bibfield{author}{\bibinfo{person}{Carl Wieman}.} \bibinfo{year}{2002}\natexlab{}.
\newblock \bibinfo{title}{PhET}.
\newblock \bibinfo{howpublished}{https://phet.colorado.edu/}.
\newblock


\bibitem[Zhang et~al\mbox{.}(2023)]%
        {zhang2023ACL}
\bibfield{author}{\bibinfo{person}{Mengxue Zhang}, \bibinfo{person}{Zichao Wang}, \bibinfo{person}{Zhichao Yang}, \bibinfo{person}{Weiqi Feng}, {and} \bibinfo{person}{Andrew Lan}.} \bibinfo{year}{2023}\natexlab{}.
\newblock \showarticletitle{Interpretable Math Word Problem Solution Generation Via Step-by-step Planning}. In \bibinfo{booktitle}{\emph{Proceedings of the 61st Annual Meeting of the Association for Computational Linguistics}}. \bibinfo{pages}{6858–--6877}.
\newblock


\end{thebibliography}

\end{document}